\documentclass[prl,showpaces,preprintnumbers,amsmath,amssymb,twocolumn,superscriptaddress]{revtex4}
\usepackage{graphicx}
\usepackage{dcolumn}
\usepackage{bm}
\usepackage{amssymb}
\usepackage{amsmath}
\usepackage{epstopdf}
\usepackage[dvipdfm, pdfstartview=FitH, CJKbookmarks=true, bookmarksnumbered=true, bookmarksopen=true, colorlinks, pdfborder=001, linkcolor=blue, anchorcolor=blue, citecolor=blue]{hyperref}
\begin{document}
\title{CeIrIn$_5$: Superconductivity on a Magnetic Instability}
\author{T. Shang}
\affiliation{Center for Correlated Matter and Department of Physics, Zhejiang University, Hangzhou, 310027, China}
\affiliation{Los Alamos National Laboratory, Los Alamos, New Mexico, 87545, USA}
\author{R. E. Baumbach}
\affiliation{Los Alamos National Laboratory, Los Alamos, New Mexico, 87545, USA}
\author{K. Gofryk}
\affiliation{Los Alamos National Laboratory, Los Alamos, New Mexico, 87545, USA}
\author{F. Ronning}
\affiliation{Los Alamos National Laboratory, Los Alamos, New Mexico, 87545, USA}
\author{Z. F. Weng}
\affiliation{Center for Correlated Matter and Department of Physics, Zhejiang University, Hangzhou, 310027, China}
\author{J. L. Zhang}
\affiliation{Center for Correlated Matter and Department of Physics, Zhejiang University, Hangzhou, 310027, China}
\author{X. Lu}
\affiliation{Center for Correlated Matter and Department of Physics, Zhejiang University, Hangzhou, 310027, China}
\author{E. D. Bauer}
\affiliation{Los Alamos National Laboratory, Los Alamos, New Mexico, 87545, USA}
\author{J. D. Thompson}
\affiliation{Los Alamos National Laboratory, Los Alamos, New Mexico, 87545, USA}
\author{H. Q. Yuan}
\email{hqyuan@zju.edu.cn}
\affiliation{Center for Correlated Matter and Department of Physics, Zhejiang University, Hangzhou, 310027, China}
\date{\today}
\begin{abstract}
We report the doping-induced antiferromagnetic state and  Fermi liquid state that are connected by a superconducting region in a series of CeIrIn$_{5-x}$Hg$_x$, CeIrIn$_{5-x}$Sn$_x$ and CeIr$_{1-x}$Pt$_x$In$_5$ single crystals. Measurements of the specific heat $C(T)$ and electrical resistivity $\rho(T)$ demonstrate that hole doping via Hg/In substitution gives rise to an antiferromagnetic ground state, but substitutions of In by Sn or Ir by Pt (electron doping) favor a paramagnetic Fermi liquid state. A cone-like non-Fermi liquid region is observed near CeIrIn$_5$, showing a diverging effective mass on the slightly Hg-doped side. The obtained temperature-doping phase diagram suggests that CeIrIn$_5$ is in proximity to an antiferromagnetic quantum critical point, and heavy fermion superconductivity in this compound is mediated by magnetic quantum fluctuations rather than by valence fluctuations.
\begin{description}
\item[PACS number(s)]
74.70.Tx, 74.62.-c, 74.40.Kb,
\end{description}
\end{abstract}
\maketitle

The heavy fermion (HF) series CeTIn$_5$ (T = Co, Rh, Ir) has provided prototype examples to study competing phases and their emergent behaviors arising from electron correlations~\cite{115reviews}. CeCoIn$_5$ is a HF superconductor with the highest superconducting transition temperature ($T_\textup{sc}$ = 2.3 K) among the Ce-based HF compounds~\cite{petrovicCo115}. A slight substitution of In with Cd or Hg tunes the system to a long range antiferromagnetic (AFM) state~\cite{gofryk2012, pham2006reversible}. Furthermore, application of pressure to the antiferromagnets CeRhIn$_5$ and CeCo(In,Cd)$_5$ eventually suppresses the AFM order and induces superconductivity (SC) near an AFM quantum critical point (QCP) (Scenario I in Fig. 1)~\cite{pham2006reversible, bauer2005, Zaum2011, TusonRh115}. The resulting superconducting phase diagram is nearly identical to that of CeCoIn$_5$ after a suitable pressure shift~\cite{sidorov2002}. Thus, it has been widely accepted that CeCoIn$_5$ sits on the threshold of a magnetic instability at ambient pressure~\cite{pham2006reversible, sidorov2002, bauer2005, Zaum2011, bianchi2003avoided,gegenwar2013}.

\begin{figure}[tbp]
     \begin{center}
     \includegraphics[width=3.4in,keepaspectratio]{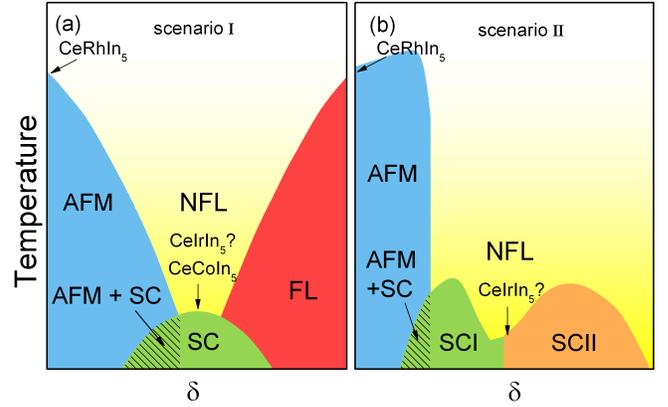}
     \end{center}
     \caption{(Color online) Schematic diagrams of two different scenarios for HF superconductors. Here $\delta$ represents a tuning parameter, like pressure or doping. (a) Scenario I: A superconducting dome appears upon suppressing the AFM transition to a QCP~\cite{Mathur1998}, which applies to most Ce-based HF superconductors including CeCoIn$_5$ and CeRhIn$_5$~\cite{pham2006reversible, sidorov2002, bauer2005, Zaum2011, TusonRh115}. (b) Scenario II: Two superconducting domes (SC I and SC II) develop upon increasing the lattice density~\cite{yuan2003observation}. SC I corresponds to Scenario I, where SC is formed via critical spin fluctuations, and SC II shows a novel superconducting state presumably arising from valence fluctuations~\cite{valence}.}
     \label{fig1}
\end{figure}

CeIrIn$_5$, a sister compound of CeCoIn$_5$, shows similar HF SC without any coexisting magnetic order~\cite{petrovic2001new}. In spite of substantial efforts to understand the exotic properties in CeIrIn$_5$, the origin of its SC still remains controversial. Resembling that of CeCu$_2$(Si,Ge)$_2$~\cite{yuan2003observation}, a scenario of two superconducting domes was proposed for CeIrIn$_5$ under combined chemical and physical pressures (Scenario II in Fig. 1)~\cite{pagliuso2001coexistence, nicklas2004, kawasaki2006}. CeIrIn$_5$ was argued to exist far from an AFM QCP and be located at a cusp-like minimum of $T_\textup{sc}$ which bridges the two superconducting domes~\cite{kawasaki2006}. Accordingly, superconductivity of CeIrIn$_5$ was proposed to be mediated by valence fluctuations rather than by spin fluctuations~\cite{kawasaki2006, kawasaki2005, yashima2012}, which are more commonly taken to mediate pairing in HF superconductors (Scenario I in Fig. 1)~\cite{yuan2003observation, valence}. However, no solid evidence of a valence instability has been revealed in CeIrIn$_5$, even though nuclear quadrupolar resonance (NQR) experiments detected some differences between the two superconducting domes~\cite{kawasaki2005, yashima2012}. On the other hand, analyses of nuclear spin-lattice relaxation experiments indicated that the quantum critical behavior of CeIrIn$_5$ is more consistent with a spin-density-wave (SDW) scenario~\cite{kambe2010}. Moreover, measurements of thermal conductivity and penetration depth support a $d_{x^2-y^2}$-type gap symmetry in CeIrIn$_5$, the same as CeCoIn$_5$~\cite{Kasahara2008, vandervelde2009, An2010}. In order to examine the above two scenarios and, therefore, to unravel the nature of SC in CeIrIn$_5$, it is crucial to establish whether it is adjacent to a magnetic or a valence instability.

In this Rapid Communication, we tune the ground state of CeIrIn$_5$ by chemical substitutions on the Ir or In sites and characterize their physical properties by specific heat and electrical resistivity measurements. It is shown that CeIrIn$_5$ lies near an AFM QCP, where non-Fermi liquid (NFL) behaviors and diverging effective mass are observed. Our results suggest that the SC of CeIrIn$_5$ is mediated by spin fluctuations, constraining a theoretical model for the origin of SC in  CeIrIn$_5$.

\begin{figure}[tbp]
     \begin{center}
     \includegraphics[width=3.4in,keepaspectratio]{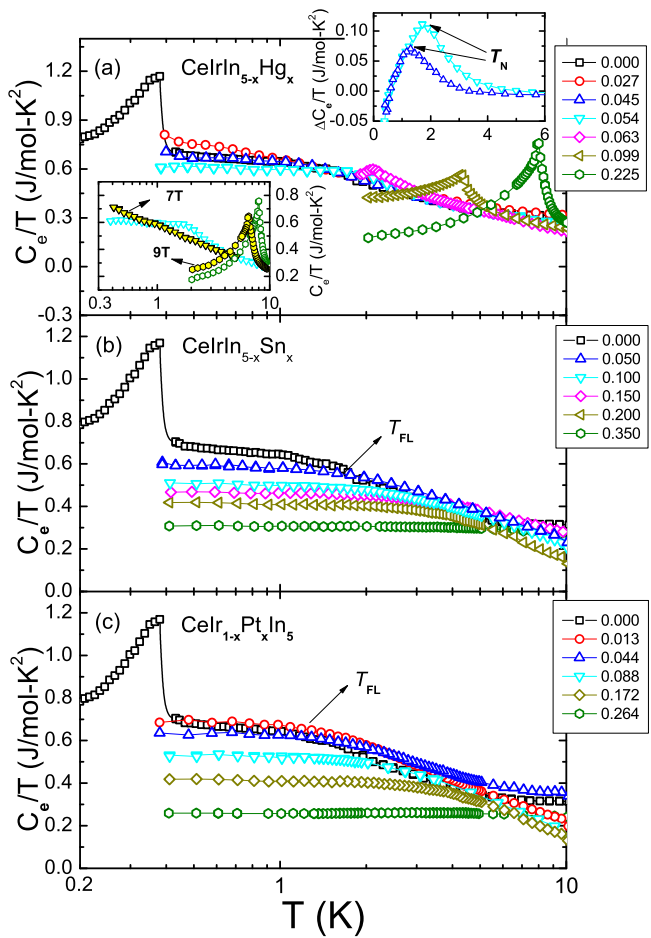}
     \end{center}
     \caption{(Color online) Temperature dependence of the specific heat for CeIrIn$_{5-x}$Hg$_x$ (a), CeIrIn$_{5-x}$Sn$_x$ (b) and CeIr$_{1-x}$Pt$_x$In$_5$ (c). The upper inset plots the specific heat $\Delta C_\textup{e}/T$ after subtracting a logarithmic contribution for $x_\textup{Hg}$ = 0.045 and 0.054. The lower inset plots the specific heat for $x_\textup{Hg}$ = 0.054 and 0.225 at a magnetic field of $B$ = 0, 7 T ($x_\textup{Hg}$ = 0.054) and 9 T ($x_\textup{Hg}$ = 0.225), respectively. The magnetic field is orientated along the $c$-axis.}
     \label{fig2}
\end{figure}

Single crystals of CeIrIn$_{5-x}$Hg$_x$ ($0 \leq x_\textup{Hg} \leq 0.225$), CeIrIn$_{5-x}$Sn$_x$ ($0 \leq x_\textup{Sn} \leq 0.35$) and CeIr$_{1-x}$Pt$_x$In$_5$ ($0 \leq x_\textup{Pt} \leq 0.264$) were grown by an indium self-flux method. Room-temperature powder X-ray diffraction confirms that all the samples crystallize in the tetragonal HoCoGa$_5$-structure. The actual concentrations of Hg, Sn and Pt were determined by microprobe analysis and single-crystal X-ray diffraction, and are $20\%$, $50\%$ and $44\%$ of their nominal values, respectively. The actual concentrations rather than the nominal ones are used hereafter. Measurements of the electrical resistivity and specific heat were performed in a Quantum Design Physical Properties Measurement System (PPMS-9T). Temperature dependence of the electrical resistivity was measured with a four-point method from 0.4 K to 300 K by a LR700 resistance bridge combined with the PPMS temperature control system.

Figure 2 shows the temperature dependence of the specific heat $C_\textup{e}/T$ on a semi-log scale for CeIrIn$_{5-x}$Hg$_x$, CeIrIn$_{5-x}$Sn$_x$ and CeIr$_{1-x}$Pt$_x$In$_5$. The electronic contributions to the specific heat $C_\textup{e}$ are obtained by subtracting a phonon contribution, with the specific heat of LaIrIn$_5$ as a reference ($C_{LaIrIn_5}/T$ = $\gamma + \beta T^2$, $\gamma \approx$ 5.8 mJ/mol-K$^2$, $\beta \approx$ 1.25 mJ/mol-K$^4$). The specific heat of pure CeIrIn$_5$ was taken from Ref.~\cite{EricCeIrIn5}, which follows NFL behavior above 1 K, i.e., $C_\textup{e}/T \sim -log T$. At lower temperatures, $C_\textup{e}/T$ tends to be a constant, which may reflect a possible Landau Fermi liquid (FL) state. The sharp jump at $T_\textup{sc}$ = 0.4 K marks the bulk SC. Upon Hg substitution, bulk superconductivity is suppressed to lower temperatures (below 0.38 K). For $x_\textup{Hg} = 0.027$, a logarithmic temperature dependence of $C_\textup{e}/T$ extends to lower temperatures. With further increasing the Hg-concentration, a kink appears at 1.3 K and 1.7 K for $x_\textup{Hg}$ = 0.045 and 0.054, respectively. Similar anomalies have been reported previously in CeCo(In,Cd)$_5$ and CuCu$_2$Si$_2$~\cite{pham2006reversible,Jeevan}, which are attributed to an AFM transition. After subtracting a logarithmic contribution of the spin fluctuations near a QCP, a pronounced transition is resolved in the specific heat $\Delta C_\textup{e}/T$ for $x_\textup{Hg}$ = 0.045 and 0.054 [see upper inset of Fig. 2(a)]. With this method, we clearly track the evolution of the N\'{e}el temperature $T_\textup{N}$ as a function of Hg-content down to very low doping. Upon applying a magnetic field, the magnetic transition is eventually suppressed. As an example, we plot the specific heat $C_\textup{e}(T)/T$ of $x_\textup{Hg}$ = 0.054 and $x_\textup{Hg}$ = 0.225 at two different magnetic fields in the lower inset of Fig. 2(a). For $x_\textup{Hg}$ = 0.054, the specific heat $C_\textup{e}/T$ demonstrates a logarithmic temperature dependence down to the lowest temperatures as the magnetic order is suppressed at 7 T. It is worth noting that the plateau below 1 K in pure CeIrIn$_5$ is very robust against external magnetic field~\cite{capan2004} and is different from the magnetic anomalies shown in $x_\textup{Hg}$ = 0.045 and 0.054. The extension of the NFL behavior and the absence of magnetic order for $x_\textup{Hg} = 0.027$ indicates that an AFM QCP lies in proximity to this Hg-content. With further increasing $x_\textup{Hg}$ ($>$ 0.054), the specific heat $C_\textup{e}/T$ shows a pronounced AFM transition: $T_\textup{N}$ increases with increasing $x_\textup{Hg}$ and reaches $T_\textup{N} = 8$ K at $x_\textup{Hg} = 0.225$. The magnetic transition is robust against magnetic field as typically seen in heavy fermion antiferromagnets, e.g., in CeRhIn$_5$~\cite{fieldHvACeRhIn5}.

\begin{figure}[bp]
     \begin{center}
     \includegraphics[width=3.4in, keepaspectratio]{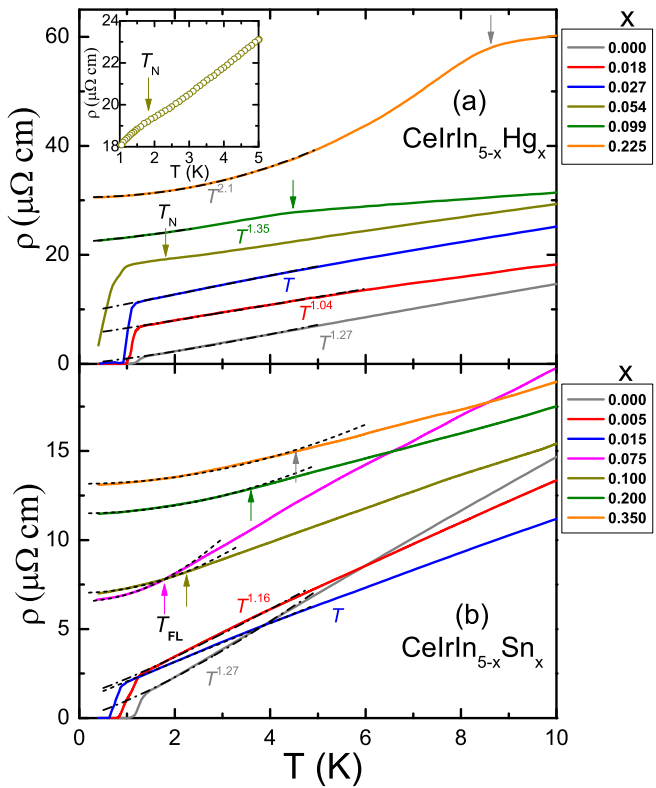}
     \end{center}
     \caption{(Color online) Temperature dependence of the electrical resistivity $\rho(T)$ for (a) CeIr$_{1-x}$Hg$_x$In$_5$ and (b) CeIrIn$_{5-x}$Sn$_x$. The inset enlarges the electrical resistivity at low temperatures for $x_\textup{Hg} = 0.054$, showing a resistive anomaly around $T_\textup{N} \simeq 1.8$ K. The arrows mark the AFM transition in the top panel and the FL temperature $T_\textup{FL}$ in the bottom panel, respectively. The dashed lines show fits to $\rho = \rho_0 + AT^2$, and the dot-dashed lines are fits to $\rho = \rho_0 + AT^n$. For $x_\textup{Hg} = 0.099$ and 0.225, the electrical resistivity is fit to temperatures below 50\% of $T_\textup{N}$.}
     \label{fig3}
\end{figure}

In contrast to Hg substitution (hole doping), electron doping via Sn/In or Pt/Ir substitutions exhibits remarkably different behaviors. In Figs. 2(b) and 2(c), we plot the specific heat $C_\textup{e}(T)/T$ for CeIrIn$_{5-x}$Sn$_x$ and CeIr$_{1-x}$Pt$_x$In$_5$. These two series of compounds behave nearly identically, being independent of the dopant sites, similar to Sn and Pt doped CeCoIn$_5$~\cite{gofryk2012}. A tiny amount of Sn or Pt dopants suppresses the superconducting transition below $T = 0.38$ K in the specific heat. No evidence of magnetic order is observed in the electron-doped compounds. Instead, the specific heat $C_\textup{e}/T$ becomes constant at low temperatures, indicating a FL ground state. The FL temperature $T_\textup{FL}$, which marks the onset temperature of the constant $C_\textup{e}/T$, increases with increasing the electron dopants. On the other hand, the specific heat Sommerfeld coefficient $\gamma$, obtained by extrapolating $C_\textup{e}/T$ to zero temperature, monotonically decreases when increasing the Sn- or Pt-concentrations.

\begin{figure}[tbp]
     \begin{center}
     \includegraphics[width=3.4in,keepaspectratio]{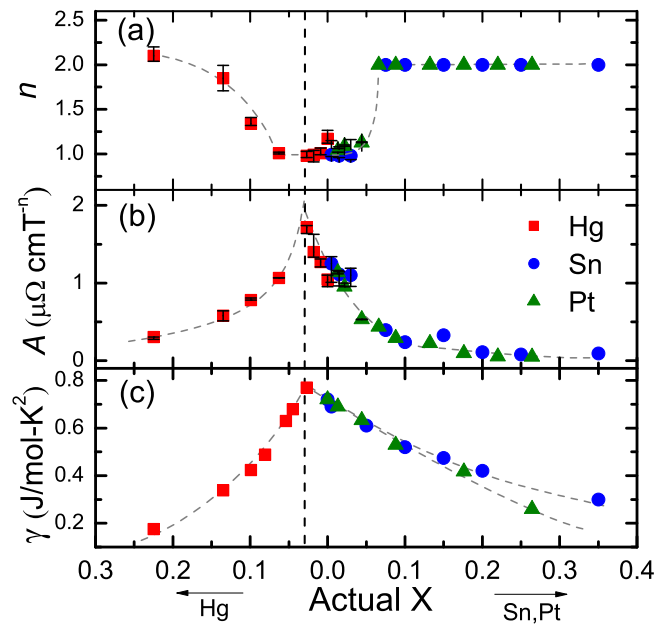}
    \end{center}
    \caption{(Color online)Doping dependence of (a) the resistive exponent $n$, (b) the $A$-coefficient and (c) the specific heat coefficient $\gamma$ for CeIrIn$_{5-x}$Hg$_x$, CeIrIn$_{5-x}$Sn$_x$ and CeIr$_{1-x}$Pt$_x$In$_5$ (To the left: Hg-content; to the right: Sn- or Pt-content). The vertical dashed line marks the critical concentration at $x^\textup{c}_{\textup{Hg}} = 0.027$. The error bars are a result of changing fit ranges.}
    \label{fig4}
\end{figure}

The above behaviors are further supported by measurements of transport properties. Figure 3 plots the electrical resistivity $\rho(T)$ of (a) CeIr$_{1-x}$Hg$_x$In$_5$ and (b) CeIrIn$_{5-x}$Sn$_x$ at several representative doping concentrations. As already seen in the specific heat data, the resistivity of CeIr$_{1-x}$Pt$_x$In$_5$ (not shown) gives nearly identical behaviors to that of CeIrIn$_{5-x}$Sn$_x$. Superconductivity only shows up near stoichiometric CeIrIn$_5$, and is suppressed by substituting either a tiny amount of Hg on the In sites or Sn on the In sites. The resistive $T_\textup{sc}$ is higher than the corresponding bulk values from the specific heat, which was argued to result from the formation of a textured superconducting phase~\cite{tusontextured}. Furthermore, for those superconducting samples their normal-state resistivity follows a behavior of $\rho =\rho_0 + AT^n$ ($n < 1.3$) at low temperatures, as illustrated in Fig. 3 by the dot-dashed lines, exhibiting a NFL behavior. The AFM transitions can be well tracked in the electrical resistivity $\rho(T)$ of Hg-doped samples ($x_\textup{Hg} \geq$ 0.045) [see the arrows in Fig. 3(a)], whose transition temperatures are highly consistent with those derived from the specific heat data. For a better illustration of the weak magnetic transition in the low-doping region, the resistivity of $x_\textup{Hg}$ = 0.054 at low temperatures is expanded in the inset, where a resistive kink marked by the arrow can be observed around $T_\textup{N} \approx 1.8$ K. With increasing the Hg-concentration, the magnetic transition becomes more pronounced and $T_\textup{N}$ shifts to higher temperatures. On the other hand, no evidence for a magnetic transition is found in the CeIrIn$_{5-x}$Sn$_x$ samples over the entire Sn-doping range. For $x_\textup{Sn} \geq$ 0.075, their low-temperature electrical resistivity can be well fitted by $\rho=\rho_0+AT^2$ [see the dashed line in Fig. 3(b)], suggesting a FL ground state as seen in the specific heat data $C_\textup{e}/T$. The FL temperature $T_\textup{FL}$, above which the resistivity deviates from the quadratic temperature dependence, increases with increasing the Sn-concentration. For all the doped compounds, the residual resistivity $\rho_0$ increases with the dopant concentration, which seems to be mainly caused by disorders introduced by elemental substitutions.

\begin{figure}[tbp]
     \begin{center}
     \includegraphics[width=3.4in,keepaspectratio]{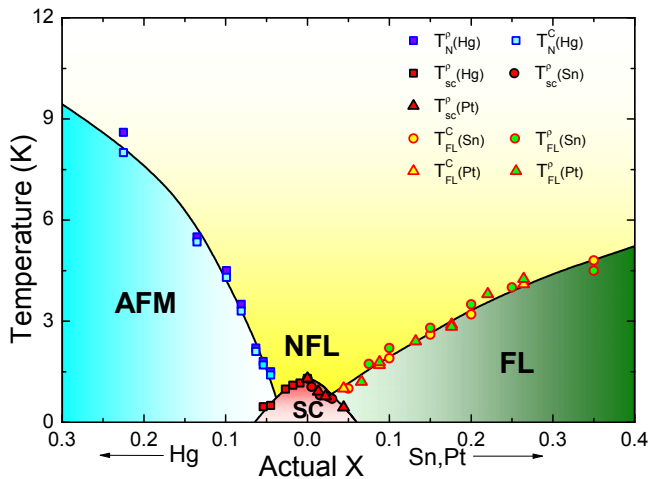}
     \end{center}
     \caption{(Color online) The magnetic and superconducting phase diagram of CeIrIn$_5$ as a function of Hg-, Sn- and Pt- doping concentration. Different symbols represent different phase transitions ($T_\textup{sc}$ or $T_\textup{N}$) or characteristic temperatures ($T_\textup{FL}$). The different colors of symbols mark different measurements of physical quantities used to determine the transition temperatures. Here the superconducting transition temperature $T_\textup{sc}^\rho$ is obtained by the mid-point of the resistive drop at $T_\textup{sc}$.}
     \label{fig5}
\end{figure}

To extend the above analyses of electrical resistivity, we fit the low-temperature resistivity with a power-law expression $\rho = \rho_0 + AT^n$ for all the measured samples. In Fig. 4, the derived parameters of the resistive exponent $n$ and the $A$-coefficient, together with the Sommerfeld coefficient $\gamma$ from the specific heat, are plotted as a function of the doping concentration for CeIrIn$_{5-x}$Hg$_x$, CeIrIn$_{5-x}$Sn$_x$ and CeIr$_{1-x}$Pt$_x$In$_5$. For $x_{\textup{Hg}} \geq$ 0.063, we simply take the lowest temperature value in $C_\textup{e}(T)/T$ as $\gamma$.
Slightly on the Hg-doped side as marked by the dashed line, both the resistive $A$-coefficient and the specific heat coefficient $\gamma$ demonstrate a diverging behavior around a critical value of $x^\textup{c}_{\textup{Hg}} = 0.027$. Furthermore, pronounced NFL behaviors with a linear-temperature dependence ($n \simeq 1$) of the electrical resistivity as well as a logarithmic-type temperature dependence in the specific heat $C_\textup{e}(T)/T$ (see Fig. 2) are observed near this critical concentration. Away from $x^\textup{c}_{\textup{Hg}} = 0.027$, a FL ground state with $n = 2$ is quickly recovered. It is noted that within the magnetically ordered state the extra electron-magnon scattering increases the resistive exponent $n$ beyond the FL prediction, e.g., $n$ = 2.1 for $x_\textup{Hg} = 0.225$.

In Fig. 5, we present a combined temperature-doping phase diagram for CeIrIn$_{5-x}$Hg$_x$, CeIrIn$_{5-x}$Sn$_x$ and CeIr$_{1-x}$Pt$_x$In$_5$, constructed from the measurements of electrical resistivity and specific heat. The left horizontal-axis stands for Hg-concentration, while the right one is for Sn- or Pt- concentration. CeIrIn$_5$ exhibits a resistive superconducting transition at $T_\textup{sc}$ = 1.2 K, but a bulk transition at $T_\textup{sc}$ = 0.4 K in the specific heat. Upon partially substituting Hg with In (hole doping), the resistive SC is observed in a narrow doping-range of $0.0 \leq x_\textup{Hg} \leq 0.054$ at temperatures above $T = 0.38$ K, and AFM order eventually develops for $x_\textup{Hg} \geq 0.045$, with $T_\textup{N}$ reaching 8 K at $x_\textup{Hg}$ = 0.225. Around the critical concentration of $x^\textup{c}_{\textup{Hg}} = 0.027$, both the specific heat $C_\textup{e}(T)/T$ and the electrical resistivity $\rho(T)$ demonstrate NFL behaviors over a wide temperature range, which can be described in terms of the spin fluctuation theory with anisotropic scattering~\cite{FH reviews}. On the other side, SC survives for the Sn/In or Pt/Ir substitution (electron doping) in the region of $0.0 \leq x_\textup{Sn} \leq 0.03$ ($0.0 \leq x_\textup{Pt} \leq0.044$). When increasing the dopant concentrations, a FL ground state is quickly recovered, as observed in both the electrical resistivity and the specific heat. The FL temperature $T_\textup{FL}$ monotonically increases with increasing the Sn or Pt concentrations. A wide cone-like NFL region sits between the AFM and FL states on top of the SC dome. These results resemble those of Cd-doped CeCoIn$_5$ and pressurized CePd$_2$Si$_2$ and CeRhIn$_5$ as schematically illustrated in Fig. 1(a) for the scenario I~\cite{pham2006reversible, sidorov2002, bauer2005, Zaum2011, TusonRh115, Mathur1998}, where SC emerges near an AFM QCP. Our results provide solid evidence that CeIrIn$_5$ is located near an AFM instability rather than a valence instability and the SC of CeIrIn$_5$ is associated with critical spin fluctuations, being similar to many other HF superconductors. We note that the strongest evidence for scenario II in CeIrIn$_5$ is due to the increase of $T_\textup{sc}$ in the pure compound under pressure~\cite{kawasaki2005, yashima2012}. However, a similar increase of $T_\textup{sc}$ under pressure is observed in pure CeCoIn$_5$ where spin-fluctuation mediated SC is generally believed~\cite{sidorov2002}. However, we cannot exclude the possibility that a degenerate valence instability may take place near the AFM QCP as theoretically argued for CeRhIn$_5$~\cite{watanable}.

In summary, we have obtained a combined temperature-doping phase diagram for CeIrIn$_{5-x}$Hg$_x$, CeIrIn$_{5-x}$Sn$_x$ and CeIr$_{1-x}$Pt$_x$In$_5$ based on a systematic study of their transport and thermodynamic properties. Upon substituting In with Hg in CeIrIn$_5$, AFM order develops at low temperatures. The system shows pronounced NFL behaviors over a wide temperature region near the critical concentration $x^\textup{c}_{\textup{Hg}} = 0.027$, where a superconducting dome is observed. Our results demonstrate that CeIrIn$_5$ lies in the vicinity of an AFM QCP and its SC is likely mediated by spin fluctuations rather than valence fluctuations. This is in line with the recent theoretical predictions~\cite{Varma2013}, and suggests a unified picture of SC in the CeTIn$_5$ family as well as many other HF superconductors.

This work is partially supported by the National Science Foundation of China (grant Nos: 11174245, 10934005, 11374257), the National Basic Research Program of China (973 Program) (2009CB929104, 2011CBA00103), Zhejiang Provincial Natural Science Foundation of China, and the Fundamental Research Funds for the Central Universities. Work at LANL was performed under the auspices of the Department of Energy, Office of Basic Energy Sciences, Division of Materials Science and Engineering.

\end{document}